# From TIGER to WST: scientific impact of four decades of developments in integral field spectroscopy

Roland Bacon
Centre de Recherche Astrophysique de Lyon

Email: roland.bacon@univ-lyon1.fr

## Abstract

This paper traces the 37 years of my career dedicated to the development of integral field spectroscopy (IFS), highlighting significant milestones and advancements. This extensive journey encompasses three generations of IFS: the initial prototype TIGER at CFHT, the first generation including OASIS at CFHT and SAURON at WHT, the second generation with MUSE at VLT, and the potential third generation represented by the Wide-field Spectroscopic Telescope (WST) project. Throughout, I discuss the lessons learned at each stage and provide my perspective on the future of IFS.

# 1. Introduction

In July 2024, I had the great honor of receiving the "Lodewijk Woltjer Lecture" award from the European Astronomical Society (EAS). The award was given "*for the development of integral field spectroscopy as a core technique in observational astrophysics and its application to a wide range of problems, in particular galaxy evolution*". This paper follows an invitation from the Astronomy and Space Science editor to write some text in connection with the award.

By design, it focuses on my own career and does not aim to be an exhaustive review paper on integral field spectroscopy. However, given that I had the chance to play a leading role in this field from the beginning, it still contains a significant, although probably biased, portion of its historical development.

What is integral field spectroscopy? It is beyond the scope of this paper to review all the details of this technique. Instead, I will provide a brief definition. For more information, I invite the reader to consult the textbook that Guy Monnet and I have written (Bacon & Monnet 2017).

We have defined integral field spectroscopy (IFS) as a subcategory of 3D spectroscopy. 3D spectroscopy encompasses all instruments that produce three-dimensional information, comprising the two spatial dimensions and one wavelength dimension. This includes integral field spectroscopy as well as other systems that obtain the third dimension through time scanning, such as the scanning Fabry-Pérot or the Fourier transform spectrometer. Integral field spectroscopy, however, produces the 3D information without time scanning. To achieve this, the input field of view is reformatted into a pseudo long-slit, which is then imaged by a classical spectrograph. There are currently three ways to perform this reformatting: the microlens concept, the fiber bundle, and the image slicer.

Often, IFS are also called integral field units, or IFUs. This terminology originally arose from developing the reformatting system to feed an existing spectrograph, with the idea that the spectrograph could then be used with different systems. Today, however, the spectrograph and the reformatting system are almost always designed together to match the two optical systems and achieve the best optical quality. Thus, in principle, these systems should be called IFS rather than IFU, but IFU is now so ingrained in common language that it is probably too late to change.

The paper is organized as follows. In the first section, "The Birth of Integral Field Spectroscopy," I describe the motivation that led us to develop this new technique and build the prototype TIGER for the Canada France Hawaï Telescope (CFHT). The first generation of IFS is then described (Section 2), with the adaptive-optics OASIS system for CFHT and then SAURON for the William Hershell Telescope (WHT), which was used to perform the first IFS survey of galaxies. I then describe the motivation for building a second generation of IFS, which led to the MUSE project for the Very Large Telescope (VLT) (Section 3). In each section, I try to give some of the lessons learned. In the last section, I give my view on the possible future roadmap for the third generation of IFS and describe the Wide-field Spectroscopic Telescope (WST) project.

## 2. The birth of integral field spectroscopy

In 1982, I began my PhD under the supervision of Guy Monnet. My research focused on modeling the stellar dynamics in elliptical galaxies. This work was motivated by the discovery that elliptical galaxies, despite being flattened, did not exhibit the expected rotation (Bertola & Capaccioli 1975). Building on Binney's (1978) pioneering work on velocity anisotropy in elliptical galaxies, I developed models to reproduce the observed kinematics of these galaxies (Bacon et al. 1983, Bacon 1985).

To compare my models with observations, I had access to a few published rotation and velocity dispersion profiles obtained along the major axis of these galaxies. These profiles were derived from long-slit spectroscopy, which was the only known method at that time to achieve spatially resolved stellar spectroscopy of extended sources. During this comparison, I encountered an initial difficulty due to the undocumented or poorly specified image quality of the observations, which prevented accurate comparison of the profiles with the model in the central regions.

To increase the sample size, Guy Monnet and I went to CFHT in May 1986 to conduct additional long-slit observations of elliptical galaxies. This was my first experience observing with a large telescope and attempting long-slit spectroscopy. I quickly realized how inefficient this method was. Not only was it tedious to properly align the slit with the major axis of the galaxy, but accurately measuring the point spread function during the observation was also highly imprecise.

This first experience convinced me that obtaining two-dimensional velocity and velocity dispersion fields would be far more effective. Such data would provide much more information and serve as a robust test of my models, which predicted interesting kinematical structures. Together with Guy, we decided to collaborate with the Marseille group, who were operating CIGALE, a scanning Fabry-Perot interferometer (Boulesteix et al. 1984). In November 1986, we returned to CFHT to conduct a test on a bright elliptical galaxy.

Although the scanning Fabry-Perot had already been successfully used to obtain velocity fields of ionized gas in extended sources (e.g. Caplan et al. 1985), it had never been tested on early-type galaxies. The challenge was that these galaxies contain little or no ionized gas, requiring the use of absorption lines to derive kinematics. With a scanning Fabry-Perot, the third wavelength dimension is obtained by time scanning, limiting the wavelength range that can be explored within a reasonable telescope time. While this limitation poses no issue when using a single bright line like H-alpha, it is challenging with absorption lines, which are broad and have much lower contrast. Additionally, the field-of-view and wavelength dependency with off-axis angles add complexity to the data reduction. We were aware of the difficulty but decided to attempt it nonetheless, given the lack of alternatives.

Alas, this test failed. The achieved signal-to-noise ratio was too low to derive any meaningful stellar velocity information. Additionally, it proved difficult to control the spectro-photometric accuracy due to the time-scanning method. As a result, even with a high signal-to-noise ratio, measuring the velocity dispersion would have been impossible.

That was the situation in November 1986 when Guy suggested discussing the problem with Georges Courtès, his former PhD supervisor and the director of the Marseille Laboratoire d'Astrophysique Spatiale (LAS). Georges Courtès had invented a concept of an optical system capable of obtaining simultaneous spectroscopic information from a two-dimensional, contiguous region of the sky (Courtès 1980). This system uses an array of micro-lenses located at the telescope's focal plane. The array produces a corresponding array of micro-pupils at its exit plane, which are then imaged and dispersed by a classical spectrograph (Figure 1). To avoid superposition of the spectra, the micro-lens array must be properly angled relative to the dispersion direction, and a broadband filter is required.

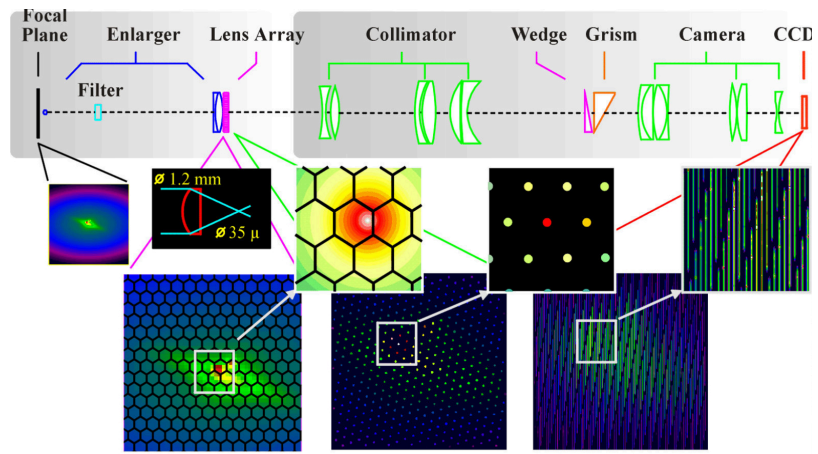

Figure 1: Concept of the micro-lens integral field spectrograph.

This concept was preferred over another approach based on a bundle of optical fibers, which are rearranged into a pseudo-slit and then sent to a classical spectrograph (Vanderriest 1980). The main reason for not opting for a fiber bundle was to avoid the throughput losses caused by the fibers' cladding, focal-ratio degradation, and low transmission in the blue. The Courtès system also had the advantage of using simple optics and eliminated the slit effect by using pupil imaging, unlike long-slit spectrographs (Bacon et al 1995). Both systems, the lens array and the fiber bundle, were christened integral field spectroscopy (Courtès 1982).

Although the fiber bundle option had been successfully tested by Christian Vanderriest at the University of Hawaii 2.2m telescope in 1980 (Vanderriest 1980), the micro-lens concept had never been tested on the sky. A brief test conducted in late 1986 during a CIGALE run at CFHT convinced us that the concept was worth developing into a dedicated prototype.

The prototype was built at the Observatoire de Marseille using an existing mechanical structure, to which we added a 271-element silica micro-lens array and a set of off-the-shelf optics (enlargers, collimator, grisms, camera) and broadband filters for the optical combination. Gérard Lelievre, the director of CFHT, was instrumental in our efforts by granting us a few discretionary nights and allowing us to use the CFHT 1024x640 RCA CCD detector.

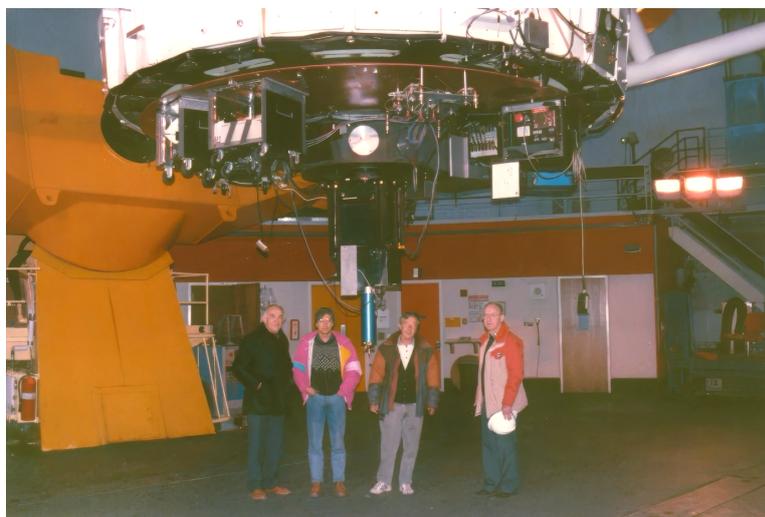

Figure 3: TIGER first light at CFHT in June 1987. From let to right: Georges Courtès, Roland Bacon, Guy Monnet and Yvon Georgelin.

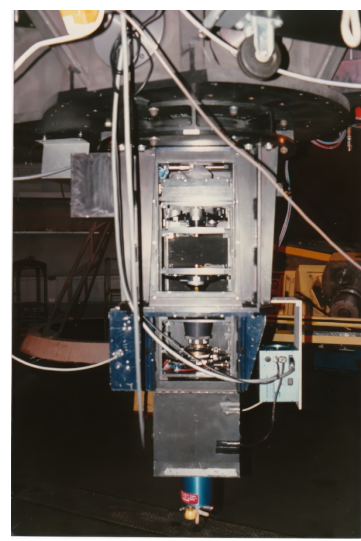

Figure 2: The TIGER prototype at the CFHT 3.60m Cassegrain focus in June 1987.

In June 1987, the integral field spectrograph TIGER, as the instrument was nicknamed, was aligned and mounted at the Cassegrain focus of the 3.6-meter telescope (Figure 2). It achieved first light shortly after (Figure 3). The first light science image of the nucleus of M51 (Figure 4) demonstrated that the system is able to provide high signal-to-noise (S/N) spectra in a relatively short exposure. This was very encouraging, and we managed to obtain enough data during this first commissioning run to quantify the system's performance and begin developing the data reduction software. The instrument and first light results were presented at several conferences (Courtès et al. 1988; Bacon et al. 1988).

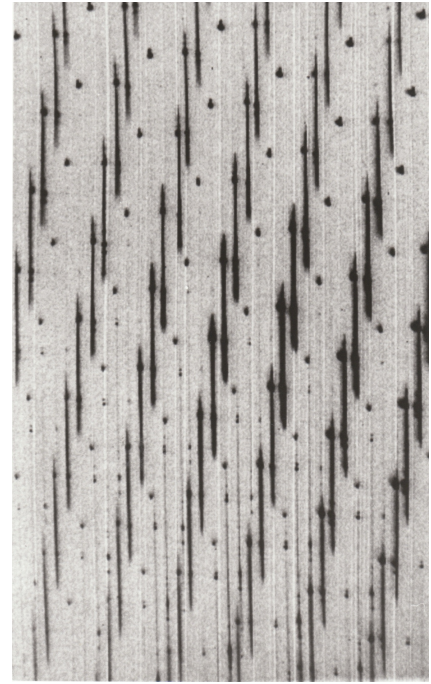

*Figure 4: This raw image of the nucleus of M51 was the first science exposure obtained by TIGER during first light at CFHT in June 1987.*

The first scientific refereed paper investigated a gravitational lens system, the Einstein Cross, and was published in 1989 (Adam et al. 1989), only two years after first light. To my knowledge, it was the first refereed science paper ever published using an integral field spectrograph.

In the few years following first light, we conducted several observing runs at CFHT and made the instrument available to other groups to broaden its scientific impact. The hardware was upgraded with a new, larger lens array of 570 micro-lenses, a 1k x 1k CCD, and a specialized optical system developed by André Baranne from the Marseille Observatory. A comprehensive refereed paper was published to present the concept and the improved prototype (Bacon et al. 1995). Developing the data reduction software was a significant effort that spanned the entire life of the instrument. Each new science project pushed the instrument's limits and required the development of dedicated observational procedures and/or data reduction recipes to meet the new scientific requirements.

In the initial years, we primarily focused on studying emission line galaxies, such as active nuclei. However, we also managed to observe the nucleus of M31 in 1990 and 1991. This resulted in the first-ever stellar velocity and velocity dispersion field, published in 1994 (Bacon et al. 1994). This achievement, coming eight years after the unsuccessful Fabry-Perot observations, opened a new frontier for the study of kinematics in early-type galaxies, which would later be further explored with SAURON.

In 1996, TIGER was decommissioned and soon replaced by OASIS. During its nine years of service at CFHT, TIGER contributed to 38 publications, including 12 refereed papers, covering a wide range of subjects: the planet Mars, young stars, nearby galaxies, active galactic nuclei, distant radio galaxies, and quasars. For more details, see Bacon et al. (1995), section 7.

These results have confirmed the validity of the concept and its ability to address a wide variety of scientific subjects. TIGER inspired the development of several instruments based on the same micro-lens concept: OASIS and SAURON, as described in the next section, as well as MPIFS at the Russian 6-meter telescope (Afanasiev & Sil'chenko 1991), SNIFS for the University of Hawaii 88-inch telescope (Aldering et al. 2002), the Kyoto 3D spectrograph operated on the Okayama 1.9-meter telescope in Japan (Ishigaki et al. 2004), and the infrared diffraction-limited IFS OSIRIS (Larkin et al. 2006) for the Keck 10-meter telescope.

In parallel to the development of TIGER, the fiber-based concept of integral field spectrograph was further implemented on 4-meter class telescopes. Notable examples include SILFID at CFHT (Vanderriest & Lemonnier 1988), DENSEPAK at KPNO (Barden & Wade 1988), and HEXAFLEX at WHT (Rasilla et al. 1990), PMAS at Calar Alto (Roth et al. 2000). The first refereed scientific paper from a fiber-based integral field spectrograph was produced with SILFID at CFHT, focusing on the study of five interacting quasar-galaxy systems (Haddad & Vanderriest 1991).

Thus, in 1996, at the time of TIGER's retirement, the concept of integral field spectroscopy was starting to gain recognition in the community. However, it was still considered a niche instrument, and long-slit spectroscopy remained the primary tool for performing spatially resolved spectroscopic observations. The slow adoption of the concept by the community can be attributed to several factors: the inherently small field of view, the complexity of data reduction, and the natural conservatism of the community. However, this was set to change with the increasing field of view enabled by larger format CCDs and the need for high spatial resolution with the advent of adaptive optics systems.

# 3. The first generation

## 3.1. OASIS

Despite being upgraded, TIGER remained a prototype and lacked the user-friendliness of common-user instruments used on 4-meter class telescopes. Each run required us to re-align the optics and re-mount the detector, which was shared with other instruments. The wheel rotation was not remotely controlled and had to be done manually. As a result, significant overheads occurred due to the frequent changes from imaging to spectrographic mode when pointing at a new object. Additionally, the mechanics had some flexure, necessitating frequent calibration during the night. In 1991, after four years of successful operation and gaining an understanding of how the prototype worked, I was convinced that it was time to build a fully new integral field spectrograph. The main reason for proposing to build an instrument for the CFHT was to open it to a broader community, rather than being limited to our collaborators.

That year, the CFHT corporation decided to launch the PUEO project, an Adaptive Optics module for the Cassegrain focus (Rigaut et al. 1998). While there was an intention to add spectrographic capability to it, no definitive plans had been made. Therefore, I proposed adding an integral field spectrograph to the system to take full advantage of the high spatial resolution provided by PUEO.

The case was straightforward to make. It was easy to demonstrate that a long-slit spectrograph becomes very inefficient when working at high spatial resolution. The first reason is that most sources display increasingly complex morphologies when viewed at high spatial resolution. While a long-slit aligned with the major axis of an elongated source observed at low spatial resolution might capture its main characteristics, this is generally not the case when spatial resolution improves. The second reason is that the spatial PSF evolution with wavelength ($\lambda$) is much more pronounced in diffraction-limited mode ($\propto \lambda^{-1}$) than in seeing mode ($\propto \lambda^{-0.2}$), making it difficult to optimize the slit width for a wide wavelength range.

The proposed instrument, named OASIS (Optically Adaptive System for Imaging Spectroscopy), incorporated an imaging mode and a TIGER mode with four different spatial samplings (0.04 – 0.30 arcsec) and a set of grisms to cover the 0.4-1 µm wavelength range. The Marseille group then proposed two additional modes: a scanning Fabry-Perot and a Pytheas mode. The latter is a combination of the TIGER mode and a scanning Fabry-Perot (Le Coarer et al. 1992). Finally, the CFHT scientific advisory committee, after reviewing our proposal, gave a positive recommendation but requested the addition of two more modes: an ARGUS fiber-bundle mode and, to my surprise, a long-slit mode. As a young PI with little experience, I did not try to argue and said yes to all these new requirements.

Today, I would be much more cautious before adding new constraints to a proposed instrument. I have learned that apparent "flexibility" often compromises the instrument's stability, and it is always better to build an instrument that performs a few tasks exceptionally well rather than many tasks poorly. A PI must have his/her own vision and sometimes should be able to take some distance with "committee recommendations" or "friends suggestions".

OASIS was built at the Lyon observatory with oversight from CFHT. While its mechanical structure was designed to support all the requested modes, limited funding and lower prioritization meant that only the imaging and TIGER modes were fully realized. Significant effort was invested to produce a state-of-the-art instrument, including a complete and documented data reduction software. OASIS achieved first light in 1997 (Figure 5), one year after the commissioning of PUEO, and was subsequently made available to the CFHT community.

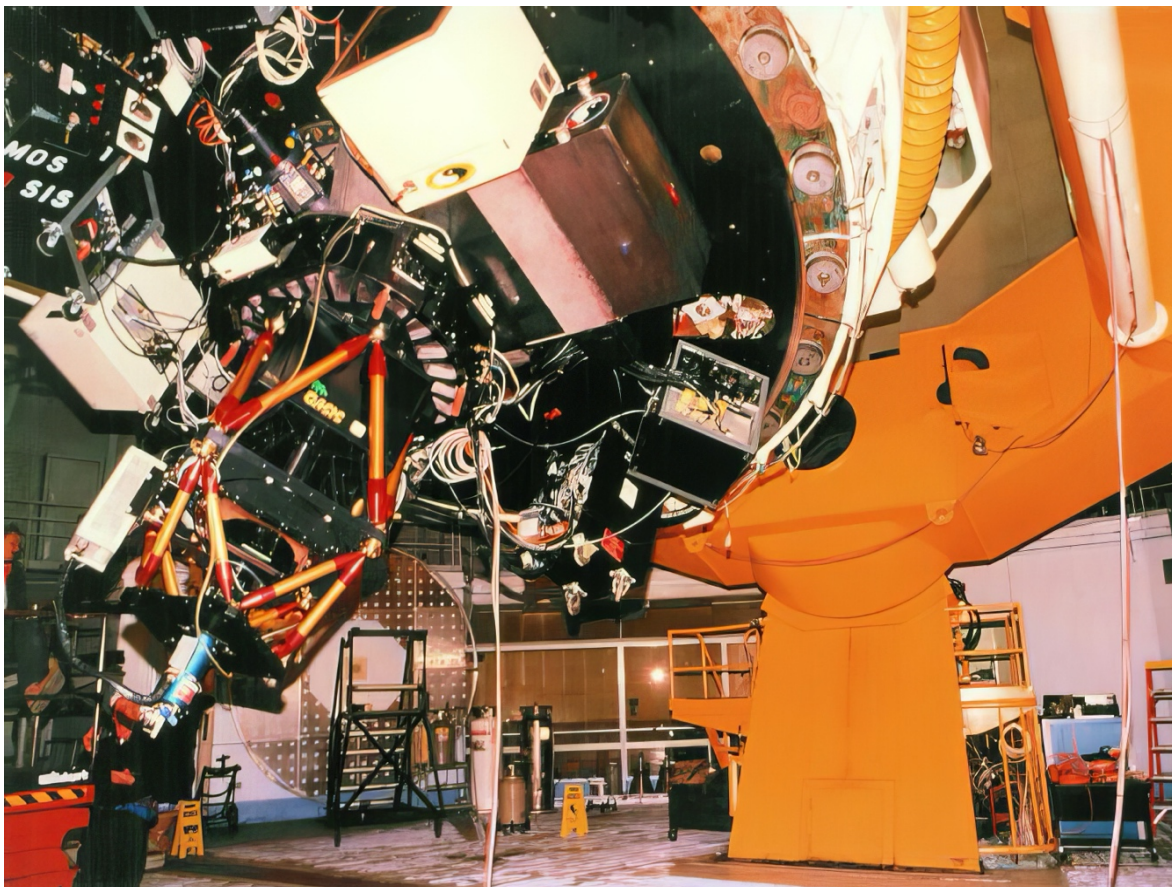

*Figure 5: OASIS at CFHT Cassegrain focus (1997).*

OASIS was operated at CFHT until 2003. Unfortunately, the instrument did not gain the expected attention and was not much requested by the community. Its scientific impact, with only eight refereed papers published over those six years, was limited. I consider this operation a failure, given the significant effort and cost invested in building the instrument.

Learning from one's mistakes is essential for progress. In my view, the reasons for this failure are manifold. The primary issue was the overly optimistic performance expectations for PUEO, the adaptive optics system. Commissioning measurements showed a two-fold improvement in FWHM in the visible range (Veran et al. 1997), but in practice, such performance was rarely achieved with OASIS. The lack of bright natural guide stars was often the culprit, and guiding on galaxy nuclei did not yield good results. The PUEO system introduced 30% throughput losses, and with the small spatial sampling, we were working in a detector noise-limited regime, which is far from ideal. Additionally, the time required to set up PUEO and close the loops was significant, leading to efficiency losses. A critical analysis of the expected PUEO performance would have revealed these issues. However, I naively assumed the PUEO's expected performance were guaranteed, despite knowing that there can be a substantial difference between a "selling" proposal and reality.

In the end, with only a marginal improvement in image quality, a small field of view, a read noise-limited regime, and a reduced number of extragalactic targets due to the lack of natural guide stars, OASIS was not very attractive to the community. Additionally, not everyone was willing to invest significant time in handling the data reduction and analysis required for this type of data.

In the meantime, CFHT decided to concentrate on wide-field imaging with the development of Megaprime. I began exploring alternative telescopes for OASIS, and an opportunity arose with the SAURON development to relocate OASIS to the 4.2-m William Herschel Telescope (WHT) on La Palma. This move was driven by the development of their own adaptive optics system, NAOMI (Myers et al. 2003), and a laser guide star called GLAS (Rutten 2004). The latter would address the lack of laser guide stars at CFHT and significantly increase the number of available targets.

After some minor adaptations, OASIS was mounted on the WHT Nasmyth platform. However, the NAOMI adaptive optics system was not yet fully commissioned, and the laser guide star project was abandoned. This delayed the effective start of OASIS operations and did not resolve its main issue of sky coverage. OASIS was operated at the WHT until 2008, producing an additional 13 refereed papers.

Despite its limited impact, OASIS was the first IFS coupled to an AO system. Future projects, focusing on the infrared where AO performance is much better, would exploit this idea: for example, SINFONI at the VLT (Eisenhauer et al. 2005), OSIRIS at Keck (Larkin et al. 2006), and NIFS at GEMINI (McGregor et al. 2007).

## 3.2. SAURON

In 1995, the OASIS project was well advanced at the Lyon Observatory. I knew that TIGER would soon be decommissioned once OASIS began operations at CFHT. The decision to build OASIS as the first-ever IFS designed for adaptive optics was driven by the opportunities presented by the development of PUEO at CFHT. However, I was convinced that significant new scientific discoveries could be made with a large field-of-view IFS, something that OASIS would not be able to achieve.

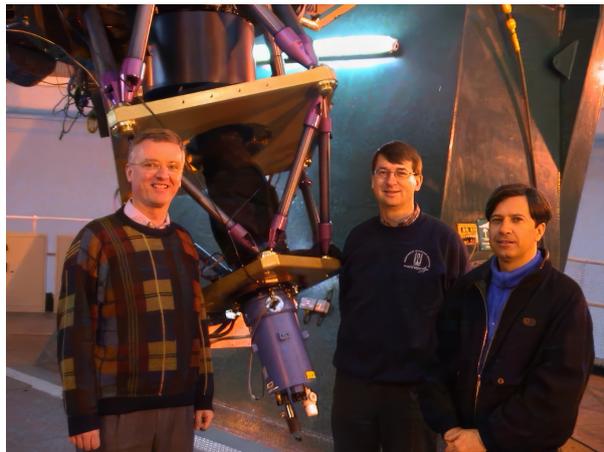

*Figure 6: The SAURON principal investigators in front of SAURON at the WHT in March 2001. From left to right : Tim de Zeeuw, Roger Davies and Roland Bacon.*

It was out of the question to propose building another IFS at CFHT, so I approached Tim de Zeeuw in Leiden to evaluate if something could be done at the 4.2-m William Herschel Telescope (WHT). The idea was to create an instrument dedicated to the study of nearby galaxies, not a common-user instrument. We approached Roger Davies in Durham, and the three of us started the SAURON project (Figure 6).

The instrument was also based on the TIGER concept, but it was designed with an entirely different philosophy than OASIS. It had the minimum number of modes and a fixed wavelength range and dispersion. The main mode of the instrument had a large field of view of 33x41 arcsec$^2$ (14 times larger than the OASIS largest field of view) and a 0.94 arcsec sampling. The throughput was optimized and the few motions were remotely controlled. It was built at the Observatoire de Lyon in less than 2.5 years, a record! It saw first light at the Cassegrain focus of the WHT in February 1999. The detail of its design and performance are given in Bacon et al. 2001.

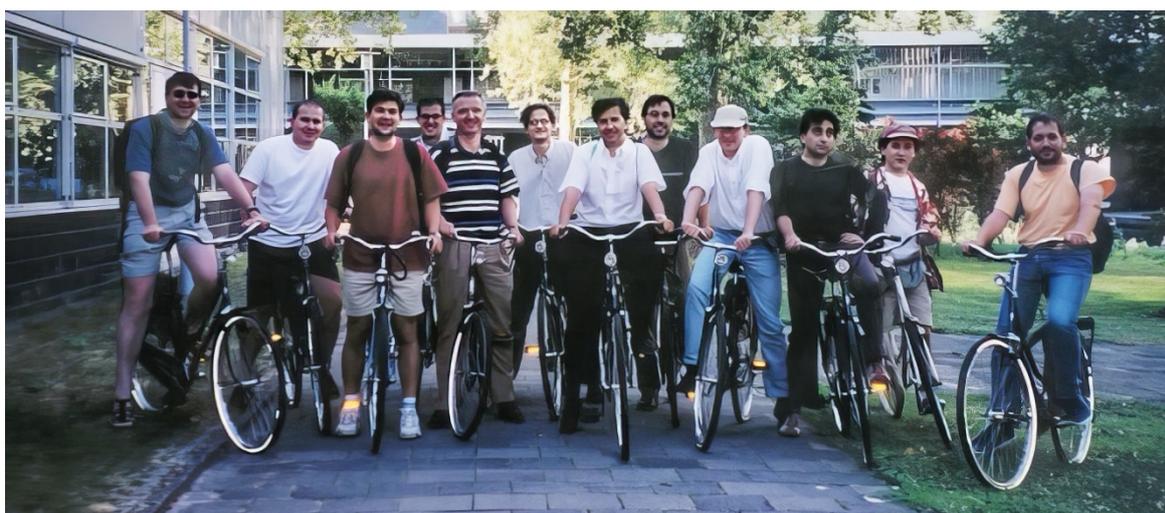

*Figure 7: The SAURON team in Leiden (July 2001). From left to right : Roger Davies, Jesus Falcon-Barroso, Davor Krajnovic, Michele Cappellari, Tim de Zeeuw, Fabien Wernli, Roland Bacon, Martin Bureau, Bryan Miller, Marc Sarzi, Yannick Copin and Eric Emsellem.*

We assembled an international team of PhDs and postdocs (Figure 7), comprising an interesting mix of instrumentalists, observers, and theoreticians. The combination of their varied expertise has been essential for conducting and scientifically leveraging the survey.

The survey aimed to study a representative sample of 72 nearby early-type galaxies (de Zeeuw et al. 2002). This sample size is an order magnitude larger than the few galaxies observed using TIGER or OASIS. Moreover, the SAURON maps were not restricted to the central regions of the galaxies but extended up to one effective radius, which was crucial for accurately characterizing their properties. Although there were no guaranteed observing nights for the survey, we managed to complete it in four years by requesting observing time from the Netherlands and UK time allocation committees.

The SAURON survey has significantly advanced our understanding of early-type galaxies. The diversity of stellar kinematics and line-strength maps (Figure 8) has shown that photometry alone cannot capture the complexity of these systems, revealing their complex internal dynamics and varied formation histories. Consequently, we introduced a new classification of these galaxies based on their kinematics: fast and slow rotators (Emsellem et al. 2007). This classification is now widely used in the study of galaxy dynamics.

The survey produced 9 PhD theses and a series of 21 peer-reviewed papers, totaling 6,250 citations (ADS Aug 2024). The most cited paper, focusing on the fundamental plane of elliptical and lenticular galaxies (Cappellari et al. 2006), reached 1,000 citations in August 2024 (ADS).

The success of this initial survey led us to propose a new initiative called Atlas$^{3D}$. This survey aimed to build on the results of the earlier SAURON survey by focusing on a complete volume-limited sample of 260 early-type galaxies within a distance of 42 Mpc. The three of us (Roger Davies, Tim de Zeeuw, and I) passed on the leadership of this new survey to four co-PIs, early-career members of the SAURON team: Michele Cappellari, Davor Krajnović, Eric Emsellem, and Richard McDermid. It is important to note that the SAURON project, in addition to providing breakthrough science, has been instrumental in training future scientists by providing a stimulating and diverse environment.

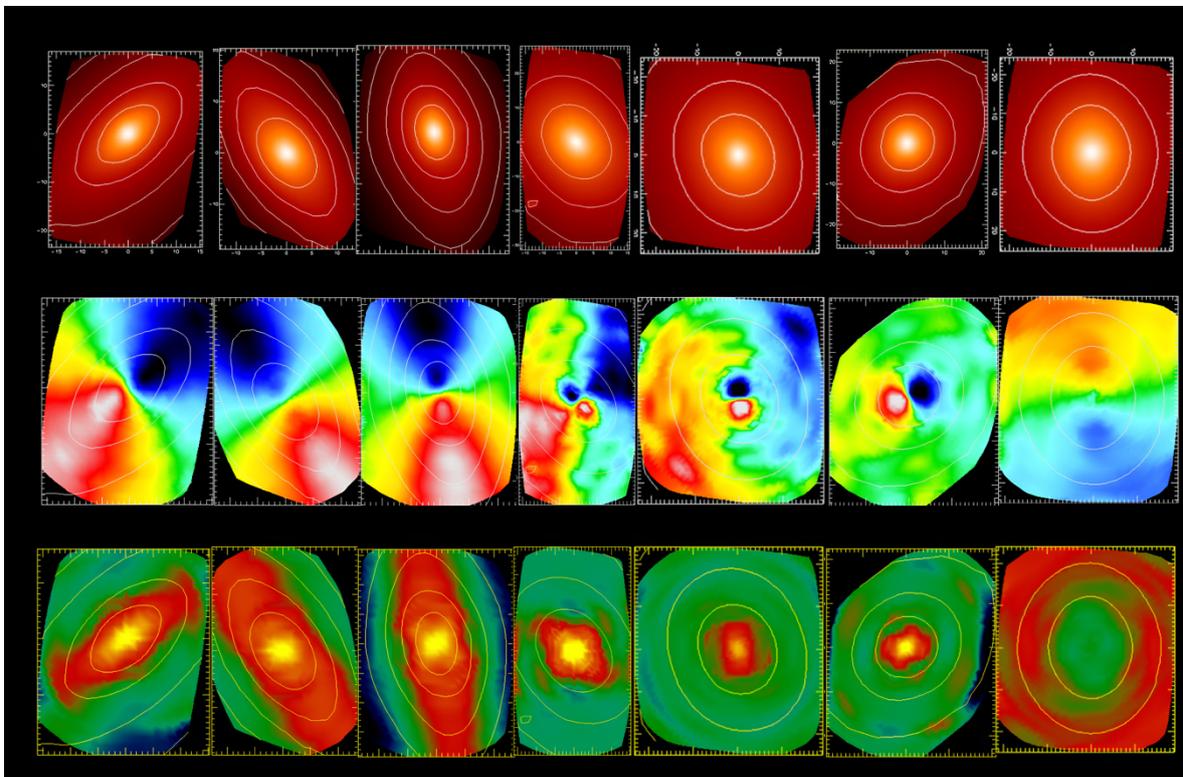

*Figure 8: Example of stellar kinematics and line-strengths maps of late-type galaxies from the SAURON survey. The top row displays the reconstructed white-light image, the middle row the stellar velocity field and the bottom row the line-strengths maps.*

The observations were carried out in 2007 and 2008. The larger sample size of Atlas$^{3D}$ facilitated the investigation of the impact of the environment and allowed for a more detailed exploration of how the fast and slow rotator classes correlate with other properties, such as environment, morphology, and stellar populations. The inclusion of multi-wavelength data provided a more detailed and comprehensive view of the evolution and diversity of these galaxies. The new survey produced 31 peer-reviewed papers, which have accumulated a total of 7,200 citations (ADS, August 2024).

SAURON surveys have been instrumental in promoting integral field spectroscopy as an essential technique for spatially resolved studies of galaxies. While earlier IFS systems such as TIGER and OASIS (lenslet-based IFS) and SILFID, DENSEPAK, and HEXAFLEX (fiber-based IFS) demonstrated the validity of the integral field spectroscopy concept, it was the scientific impact of the SAURON surveys that convinced the community of the power of this new technique.

Inspired by the SAURON surveys, the community has launched several major surveys of both nearby and distant galaxies: e.g., SINS on the VLT (Forster Schreiber et al. 2009), CALIFA on Calar Alto (Sanchez et al. 2012), MaNGA as part of the Sloan Digital Sky Survey (Bundy et al. 2015), and SAMI on the AAO (Bryant et al. 2015). Together, these efforts and their scientific results have established IFS as a major tool for astronomy, marking the end of long-slit spectroscopy.

## 4. The second generation

In 2001, ESO issued a call for ideas for the second generation VLT instruments. At that time, we were in the midst of the SAURON survey, with five successful runs already completed at the WHT. We were quite busy with significant amounts of data to digest, analyze, and publish. Nevertheless, such an opportunity to build a new IFS for the VLT could not be missed. With the growing scientific impact of SAURON, I was convinced that we could make a strong proposal.

### 4.1. MUSE

The easy solution would have been to propose an extension of SAURON adapted to the VLT with improved performance. However, I was aiming for something far more ambitious. My idea was to design a "true 3D instrument" that combines the expected performance of an imager — large field of view, high spatial resolution, and excellent throughput — with the best capabilities of modern spectrographs: a wide spectral range and good spectral resolution.

These top-level requirements are, of course, valid for any integral field spectrograph, but, for practical reasons, they were never achieved in the first generation of IFS. The main challenge is integrating all these requirements: a large field of view and high spatial resolution, plus a wide spectral range and good spectral resolution, implies a very large number of detector pixels. This leads to enormous optics, resulting in high costs and poor throughput. For this reason, the first generation IFS were specialized either in high spatial resolution, e.g., OASIS or SINFONI (Eisenhauer et al. 2003), but with a correspondingly small field of view, or in wide field like SAURON but with poor spatial resolution.

This is the primary difficulty in achieving imager performance, but there is a second, more subtle issue intrinsic to the IFS concept. In an IFS, the spectrographic capabilities are provided by a classical spectrograph, i.e., a combination of a collimator, a disperser, and a camera. Therefore, IFS spectrographic performance is generally as good as any long-slit spectrograph. However, for imaging capabilities, the reformatting of the input field of view into a pseudo-slit, regardless of the technology used (lenslet, fibers, or slicers), is much more complex. Image reconstruction needs to account for all the imperfections of the various optical elements involved in this reformatting. This is not an easy task and may explain why the imaging capabilities of the first generation of IFS were not at the level expected for an imager. Throughput is another challenge. Compared to a classical imager, the number of optical elements in the light path is significantly larger, which can lead to substantial throughput losses if the design, coatings, and detector performance are not optimized and carefully monitored during manufacturing.

An extension of the current IFS design was impractical to meet these challenges. A paradigm shift was required, leading to the concept of multi-units. In this new scheme, rather than building a huge optical system, you split it into a number of identical modules. Each module is then easier to manufacture, and if you have enough of them, you can reduce the overall cost through the economies of scale.

The question of which IFS technology was best suited for such an instrument was heavily debated. Lenslet and fiber systems each have their pros and cons. The TIGER concept had the merit of optical simplicity and higher throughput, but it was limited to a small wavelength range due to the specific spectral arrangement on the detector. On the other hand, the fiber system made much better use of pixel space in the wavelength direction but suffered from low throughput, especially at blue wavelengths. The best packing efficiency, i.e., the ratio of the total number of detector pixels to the number of used voxels (spatial and spectral pixels), is achieved by the slicer technology with 90%, while the fiber and lenslet systems achieve only 75% and 50% packing efficiency, respectively.

The image slicer, initially based on flat mirrors, was first used in SINFONI (Tecza M. et al. 2000). However, for a large field of view, flat mirrors result in overly large spectrograph optics. To address this issue, we proposed using the advanced slicer concept (Content 1997), which relies on off-axis very thin spherical mirrors. The fabrication of these mirrors posed a significant challenge and potential risk for the instrument, necessitating the development of prototypes.

I assembled a consortium with the necessary expertise, initially composed of the following six institutes: AIP, CRAL, ETH, LAM, the Durham Physics Department, and Sterrewacht Leiden. This composition evolved later due to national funding constraints or staff availability. The final consortium composition was established in 2004 with the following institutes: AIP, CRAL (lead), ETH, Göttingen University, IRAP, and Sterrewacht Leiden. ESO was also technically involved in the project, taking responsibility for the detector system.

After some iterations, we converged on an instrument design based on 24 identical modules, each composed of an image slicer, a spectrograph, and a 4k x 4k detector. This configuration translates to a field of view of 1x1 arcmin² with a spaxel sampling of 0.2 arcsec, a simultaneous wavelength range of 465 to 935 nm, and a spectral resolution of 3000 in the middle of the band. The instrument was christened MUSE, short for Multi Unit Spectroscopic Explorer.

In addition, the ESO adaptive optics group was advocating for the development of an adaptive optics facility (AOF, Arsenault et al. 2004). The aim of the AOF was to provide high spatial resolution to the VLT Nasmyth focal station using a deformable secondary and four laser guide stars. Two applications were foreseen: a laser tomography adaptive optics (LTAO) mode providing diffraction-limited image quality over a small field of view (a few arcseconds) and a ground-layer adaptive optics (GLAO) module providing improved image quality over a large field of view (an arcminute). Thanks to the laser guide stars, the sky coverage would be extensive, addressing one of the main limitations of OASIS. Additionally, the use of an adaptive secondary would not introduce any additional optical system, thus maintaining high throughput.

The combination of MUSE and the AOF was ideal, and thus we proposed MUSE as one of the two instruments to benefit from it. To take advantage of the LTAO diffraction-limited regime, we added a second mode to MUSE, the Narrow Field Mode (NFM), by introducing another simple optical system in the foreoptics to magnify by a factor of 8, providing a sampling of 25 milli-arcseconds and a field of view of 7.5x7.5 arcseconds². The main mode was called WFM for wide-field mode.

Such an instrument would obviously have a strong impact on the study of nearby galaxies. It would offer a larger field of view and a wider spectral range than SAURON, better image quality than OASIS, and four times the collecting area of a 4m class telescope. A science case for this was easy to justify. But I had an additional science case that I wanted to push forward: the study of the high-redshift Universe by the mean of 3D deep-fields.

At that time, ground-based spectroscopic observations of distant galaxies were exclusively performed with multi-object spectroscopy (MOS): e.g., VIMOS at VLT, DEIMOS at Keck, GMOS at Gemini, and AAOmega & 2dF at AAT. I was convinced that an IFS like MUSE could be very complementary to MOS studies. One might wonder how an IFS can compete with a MOS, given that even with the field of view foreseen for MUSE (1 arcmin²), it is hundreds of times smaller than any equivalent MOS field of view. For example, DEIMOS and VIMOS have fields of view of 16.7x5 arcmin² and 14x16 arcmin², respectively.

The rationale is that MOS observations are in practice inefficient for dense fields. There is always a minimum distance between fibers, which prevents addressing fields with densities larger than a few per arcmin². At high redshift, because of cosmological dimming, one needs to go deeper to observe the population of galaxies, especially if we want to observe the low mass population. With a long integration (e.g. 80 hours), we computed that we should be able to detect hundreds of galaxies in the one arcmin² field of view of MUSE, most of them being at high redshift.

The second, more fundamental, reason, is that MOS observations, by construction, require source pre-selection. In most cases, colors are used, such as the Lyman-break pre-selection. This introduces a strong bias and significantly limits the discovery space. An IFS has no pre-selection and will obtain spectroscopic data for everything in its field of view. In particular, galaxies that are below the detection limit in broad-band imaging will never be selected for MOS observations, but they might be detected by MUSE if they have high equivalent width emission lines. Our simulations predicted that a fraction of high equivalent width Lyman-alpha emitters without counterparts in the deep broad-band images might be detected by MUSE.

In February 2002, we delivered our pre-phase A document to ESO. The proposal was positively received, and we were encouraged to continue to phase A (i.e. conceptual design and feasibility study). Over the following two years, we developed the instrument concept, refining the top-level and system requirements and the science case. We also prototyped the image slicer, which was identified as the main technical risk, and worked on obtaining a credible cost and planning estimate.

In April 2004, the concept design review took place at ESO Garching. The science case was reviewed, together with technical documentation and the management plan. The science referees were quite supportive of the nearby galaxy science case but were much less convinced by our high-z proposal. Their main objection was that MUSE wouldn't be competitive with MOS because it wouldn't have deeper capabilities and would have a much smaller field of view, where most of the spaxels would just capture the sky. This skepticism was probably shared by a large fraction of the high-z community, and MUSE, until its first light, was seen as a perfect machine for the study of extended nearby objects, but not more.

Overall, the review conclusions were positive, and the project was accepted with a number of urgent action items to be resolved before the preliminary design review. The contract was approved by the ESO Council in the summer of 2004, and we proceeded to the next phases. This marked the beginning of a very long process that spanned seven years.

Although the consortium had substantial experience in instrumentation, the study and manufacturing of 24 identical modules — each nearly the size of a first-generation IFS — posed a significant challenge. Today, with the advent of the ELT's massive and complex instrumentation program, the MUSE project might seem straightforward. However, in 2004, none of us, including ESO, were fully prepared for such an endeavor. We had to plan and manage this effort all together with great care.

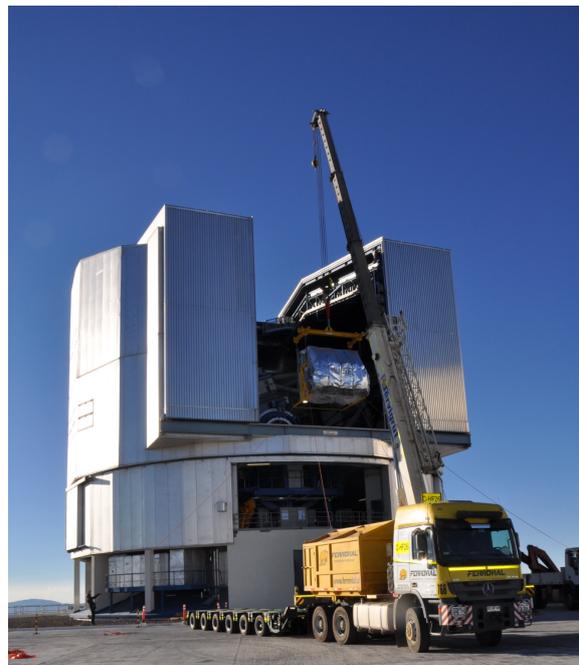

In September 2013, after a two-year delay from the original phase A schedule, ESO granted the Preliminary Acceptance in Europe (PAE), and the instrument was sent to Paranal. At Paranal, it was re-assembled in the integration hall at the base camp before being transported up to the summit. On January 19, 2014, the instrument was lifted 20 meters above the ground and literally landed on the UT4 Nasmyth platform (Figure 9) — a truly nerve-wracking moment. MUSE is a

*Figure 9: MUSE "landed" on the UT4 Nasmyth platform on January 19, 2014.*

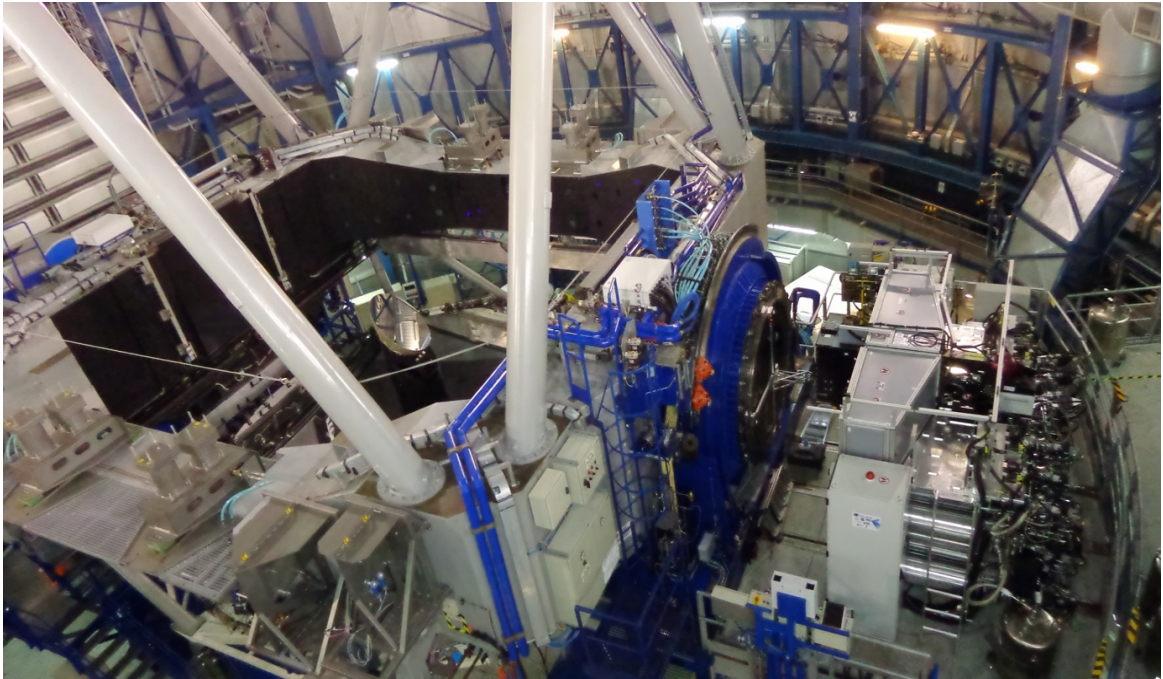
*Figure 10: MUSE positioned at the Nasmyth focal plane station of UT4.*

large instrument that almost entirely fills the Nasmyth platform, which had to be extended to accommodate it (Figure 10).

MUSE's first light was achieved soon after, on January 31, 2014 (Figure 11). Following a few commissioning runs, it was put into operation and opened to the community in October of the same year. The AOF system was commissioned a few years later, in 2017 for the GLAO mode and in 2018 for the LTAO mode. Since then, adaptive optics has been the regular mode of operation for MUSE (Figure 12).

By all standards, MUSE has been a major success. Not only is it the most in-demand instrument of the VLT, with an oversubscription factor of more than 10, but it is also the most productive, generating 200 refereed papers in 2024, accounting for 25% of the total VLT scientific production (Figure 13).

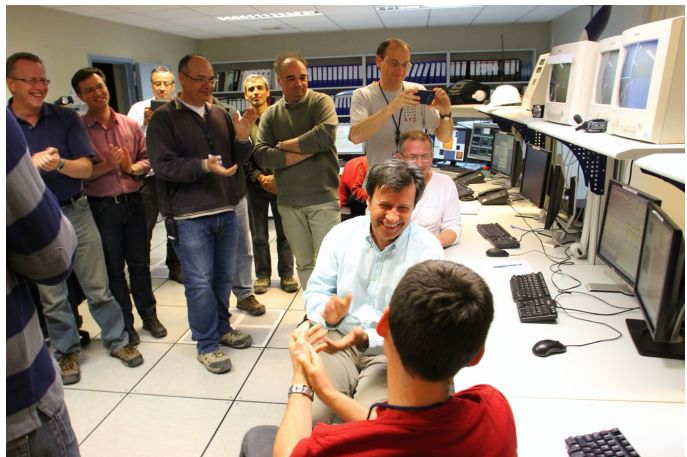
*Figure 11: MUSE achieved first light at Paranal on January 31, 2014.*

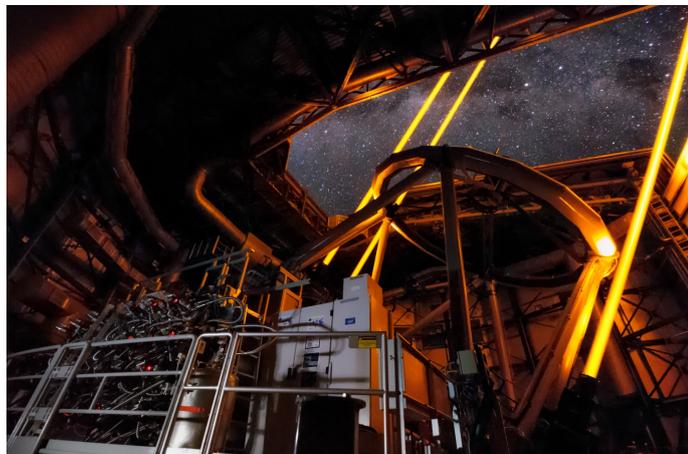
*Figure 12: MUSE with the four laser guide stars of the Adaptive Optics Facility (AOF) in operation.*

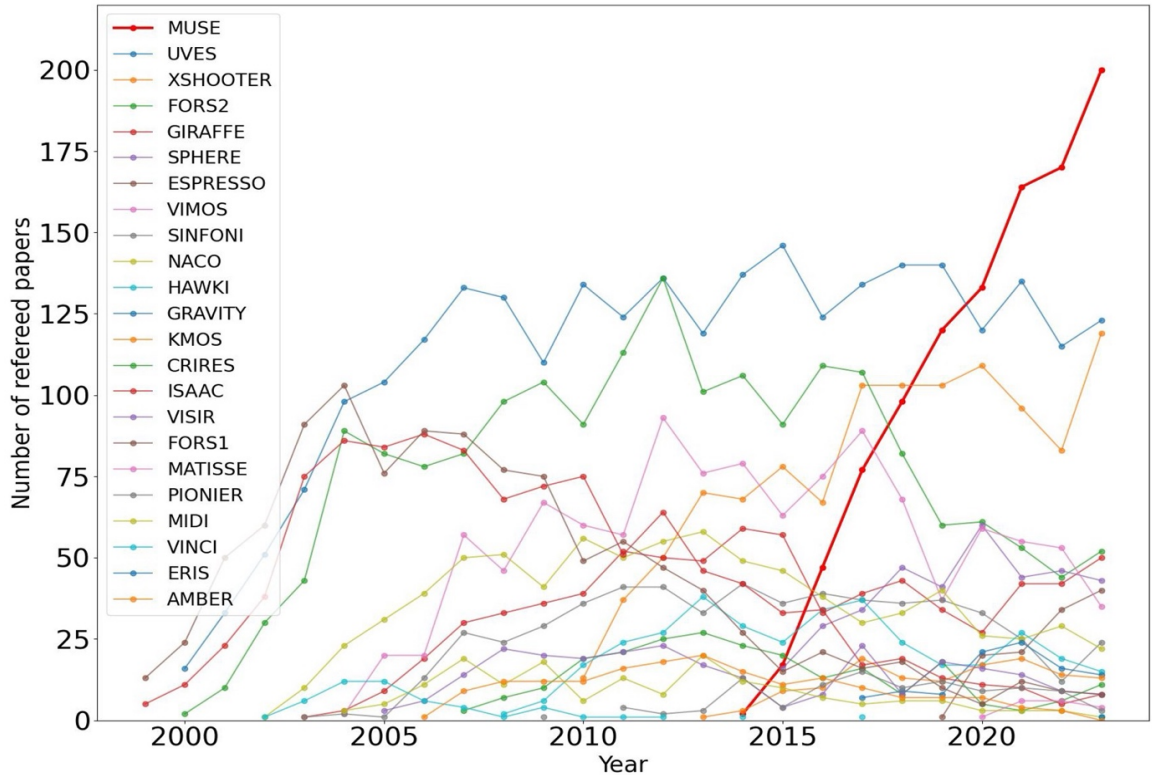

*Figure 13: Annual scientific production of MUSE compared to other VLT instruments (Source: ESO telbib).*

It is beyond the scope of this paper to summarize the main scientific results achieved with MUSE. For a detailed overview of the major highlights obtained by the MUSE consortium, I refer the reader to the publication by Bacon et al. (2024). Note that this represents only a small fraction of all MUSE publications and reflects the specific scientific interests of the consortium.

The reasons for MUSE's success are multiple. The performance of the instrument, of course, is the primary factor. Looking back at the initial goal of building a truly 3D imager, I can confidently say that we have made significant strides in achieving it. The large field of view, wide simultaneous spectral range, and good spectral resolution were integral to the construction. Image quality was ensured first through the optimization of the optical design, manufacturing, and alignment of the system, and subsequently by the AOF, which has been a fantastic achievement. The throughput was another major accomplishment, achieved through an innovative slicer and spectrograph design, as well as the use of state-of-the-art mirrors and lens coatings. Last, but not least, the advanced data reduction system has been crucial, producing high-quality, science-ready datacubes.

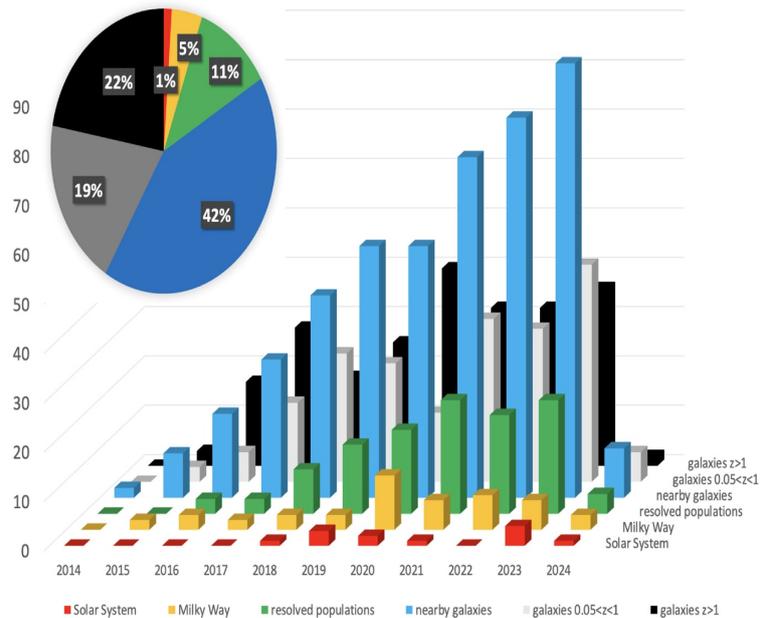

*Figure 14: Repartition of MUSE publications by source categories (from Roth 2024).*

MUSE's ability to deliver new scientific results across a wide range of subjects and targets is another reason for its success. In a recent publication (Roth 2024), the distribution of MUSE publications by source categories was studied (Figure 14). As expected, nearby galaxies represent a major subject (42%), but distant galaxies are also significant, accounting for 19% and 22% of the publications for intermediate ($0.05 < z < 1$) and high redshift ($z > 1$) populations, respectively. Resolved stellar populations in dense stellar systems within the Milky Way and the local group contribute to 11% of the publications, while the remainder is split between the Milky Way (5%) and the solar system (1%).

It is satisfying to see that, with 22% of the publications, MUSE has clearly made an impact on the field of high-z galaxies. The 3D deep fields envisioned at the start of the project have been very successful, providing a wealth of information on the Lyman-alpha population at z=3-7, including the detection of a large population of high equivalent width LAEs without counterparts in the Hubble Ultra Deep Field (Bacon et al. 2023). However, the most significant results, which were not anticipated in our phase A science case, include the ubiquitous detection of Lyman-alpha halos at z>3, allowing us to trace the circum-galactic medium in emission at high redshifts (Wisotzki et al. 2018).

This example highlights another reason for MUSE's success: its discovery power. The concept of obtaining spectroscopy of everything within a field of view has led to many unexpected discoveries. This is a fundamental feature of a game-changing facility. The number of publications that rely solely on data from the MUSE ESO archive is a testament to this.

I would like to offer my own perspective on how we collectively managed to achieve such tremendous success. Learning from the success of SAURON and, by contrast, the failure of OASIS, I set up the project to have the minimum number of modes, specifically ensuring there were no moving parts in the 24 modules. Because of the size and complexity of the system, I established stringent requirements for reliability. Essentially, while MUSE was difficult to build, it was designed to be easy to operate. Efficiency was also a key requirement, both for MUSE itself and for the AOF system. This meant aiming for the best possible throughput and maximizing open-shutter time by minimizing overheads.

Given the volume and complexity of the data, I recognized that it was essential to invest significant effort in data reduction, not just up to first light but continuing through the scientific GTO exploitation. Understanding an instrument thoroughly cannot be separated from its scientific application, which takes time. In this regard, the ESO GTO system, where the consortium receives GTO time in exchange for the technical staff's effort in building the instrument, is a win-win solution to facilitate this process. The consortium secured substantial external funding (e.g., four ERC grants) that enabled not only the scientific exploitation but also the development of the data reduction and analysis system, which was subsequently made available to the community.

But from my perspective, the most essential factor was the strong and enjoyable collaborative spirit within the consortium that we managed to build and maintain throughout the project's life. This includes our positive and trustful relationship with ESO. Given the importance of industrial contracts, establishing an early and good rapport with our industrial partners was also crucial. In most cases, we succeeded in creating a truly collaborative spirit that went beyond the typical client-supplier relationship.

Finally, I had a clear vision of the technical and scientific goals of MUSE. This vision was crucial for making tradeoffs and keeping the project on track despite the inevitable difficulties that come with such a long and challenging endeavor.

## 4.2. BlueMUSE

MUSE is quite unique. The closest instrument is KCWI, which was put into operation in 2018 at Keck (Morrissey et al. 2018). However, it is not a real challenger, with a 20 times smaller field of view when used with its 0.35 arcsec sampling, a smaller simultaneous spectral range, and the lack of adaptive optics. There is one characteristic, however, that MUSE lacks: its shortest wavelength is 480 nm, whereas KCWI covers the blue wavelengths down to 350 nm. This blue sensitivity, combined with its northern location, still makes KCWI complementary to MUSE.

With the tremendous success of MUSE and the consequent high oversubscription rate, it has become clear that there is a need for another MUSE-like instrument on the planet. After considering various options, we concluded that the most efficient approach, given ESO and the community's experience, would be to design a new MUSE for another VLT unit. However, rather than simply duplicating MUSE, we opted to design an instrument to cover the bluer spectral window (350-580 nm) with increased spectral resolution (R~3500) with respect to MUSE.

The instrument is called BlueMUSE to emphasize its affiliation. Apart from the blue wavelength coverage, BlueMUSE has the same main characteristics as MUSE, notably a field of view of 1x1 arcmin². Given the expected poor performance of adaptive optics at those wavelengths, BlueMUSE is a seeing-limited instrument and therefore operates in a single mode. One key element is to achieve the best possible throughput at blue wavelengths (Figure 15): the project aims to reach an end-to-end throughput at zenith of 15% at 350 nm and 33% at 450 nm (including the VLT and atmosphere). The blue spectral coverage and higher spectral resolution open new scientific opportunities, which have been detailed by Richard et al. 2019.

A consortium of nine institutes was set up: CRAL (lead), CEA, Astralis, AIP, CAUP, EPFL, Geneva Observatory, Stockholm University, and Durham University. The project is led by Johan Richard from CRAL. The proposal was recommended by ESO committees in 2020 for a phase A start in 2022. However, delays occurred due to the pandemic and a lack of resources at ESO, and the project's official phase A finally started in early 2024. First light is expected at the VLT in 2031.

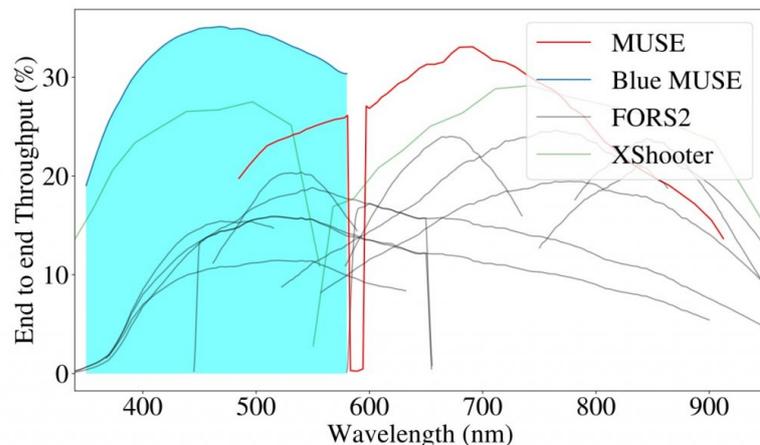

*Figure 15: Comparison of the BlueMUSE (predicted) and MUSE (measured) end-to-end transmissions with other VLT instruments. A 15% is included for slit spectrographs to account for slit losses (from Richard et al. 2019).*

# 5. Next Steps

Integral field spectroscopy is now well established in the landscape of astronomical instrumentation, having definitively replaced long-slit spectrographs for resolved spectroscopic studies. Every large telescope has at least one IFS. For the future IFS roadmap, one can envision three distinct paths: AO-assisted IFS for high spatial resolution work, the development of space-based IFS, and wide-field IFS facilities.

With the development of adaptive optics, there will be a growing need for IFS adapted to high spatial resolution. For example, on the VLT, there are ERIS, the recently commissioned near-IR IFS (Davies et al. 2018), and MAVIS, the future visible MCAO IFS (Rigaut et al. 2020). The upcoming or planned extremely large telescopes have all incorporated such IFS in their instrumental suite: HARMONI for the E-ELT (Thatte et al. 2016), IRIS for the TMT (Larkin et al. 2016), and GMTIFS for the GMT (McGregor et al. 2012). For feasibility reasons, most of these new instruments focus on the near-IR, but the advent of new AO systems optimized to work in the visible, such as the AOF on the VLT, will likely be extended to other 10m class telescopes. They have the advantage of providing higher spatial resolution than in the near-IR, and the IFS are easier and less expensive to build (no need of cooling and avalability of affordable detectors).

With two IFS instruments on JWST, namely the NIRSpec-IFU (Arribas et al. 2007) and the MIRI IFS (Wright et al. 2015), integral field spectroscopy has established its place on space telescopes. However, it is still less developed than in ground-based astronomy. To my knowledge, except for Athena with its X-IFU (Barret et al. 2016), none of the planned space telescopes (e.g., Roman, LUVOIR, OST) foresee an IFS in their instrumentation. One reason is that IFS requires a lot of detector space and large optics, which increases the instrument's mass budget. For many applications, space-based instruments prefer to use slitless spectroscopy, which provides a large field of view with much simpler optical system and minimal detector space. However, this comes at the cost of lower spectral resolution and complex operational challenges to solve the problem of source contamination. The development of more compact IFS systems may help improve the attractiveness of IFS in space instrumentation.

As demonstrated by MUSE's success, wide-field IFS on large telescopes are scientifically very attractive. It would be highly desirable to achieve a much larger field of view than the current MUSE (and future BlueMUSE) 1x1 arcmin² field of view, while keeping the other characteristics intact (e.g., spatial and spectral resolution, simultaneous spectral range). With a larger field of view, it becomes possible to survey larger regions of the sky, increasing the potential for discoveries.

However, as seen in Figure 10, MUSE completely fills the VLT Nasmyth platform. To achieve something much larger, we need to change our paradigm in the way we design and build IFS, similar to the shift from the first to the second generation of IFS. This time, we cannot design the instrument independently of the telescope; we need to design and build an integrated facility. This is the only way to advance to the third generation of IFS.

Now that ESO and the community have begun to consider the next large facility for the post-ELT era, the time is right to launch this new idea. For the past two years, a large consortium of European and Australian institutes has been working on the concept of such a facility: WST, the Wide-field Spectroscopic Survey Telescope (Bacon et al. 2024). The consortium has recently been funded by the European Horizon program to advance this concept study over the next three years.

WST is a facility consisting of a 12-meter telescope with a very wide field of view and two instruments: a MOS with low and high spectral resolution and a wide-field IFS. As shown in Figure 16 the 13-arcmin diameter patrol area of the IFS is located in the center of the 2-degree MOS field of view. The MOS low- and high-resolution modes have a multiplex of 20,000 and 2,000 and spectral resolutions of ~3,000 and ~40,000, respectively. The IFS has a field of view of 3x3 arcmin² and a spectral resolution of ~3,500. The MOS low-resolution mode and the IFS simultaneously cover the full optical range from 370 to 970 nm, while the MOS high-resolution mode will have three to four spectral windows. MOS and IFS instruments are operated simultaneously, increasing the overall efficiency and flexibility of the system. For example, while the MOS is reconfigured to match source clustering, the IFS can mosaic to cover a wider field of view, up to 9x9 arcmin².

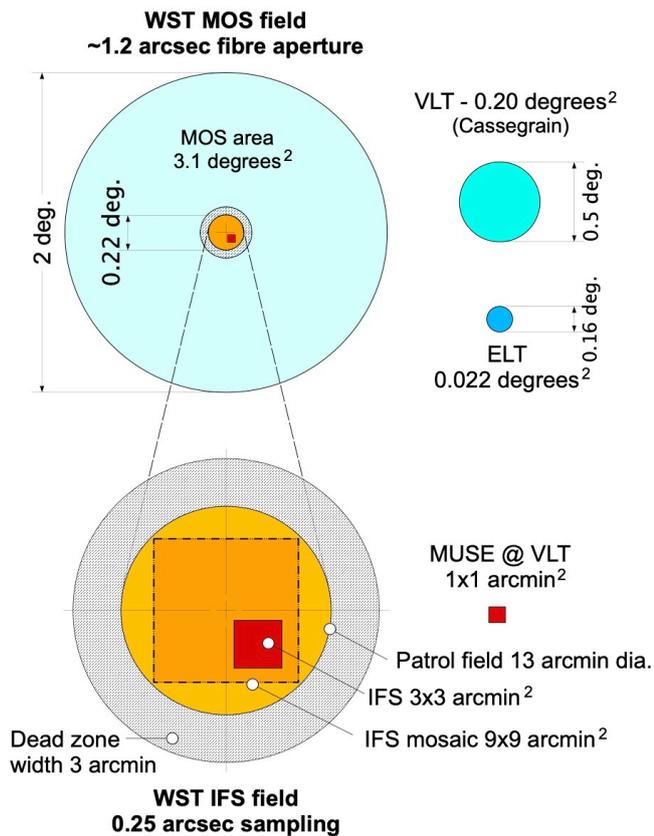

*Figure 16: WST fields of view (FoV). The top panel shows the MOS FoV and the central circular area available for IFS observations. The bottom panel offers a closer view of the latter. The IFS 3x3 arcmin² FoV (in red) can be moved within the available area, providing the 9x9 arcmin² mosaic capability A distinct dead zone separates the MOS and IFS FoVs. The VLT, ELT and MUSE FoV are represented for comparison.*

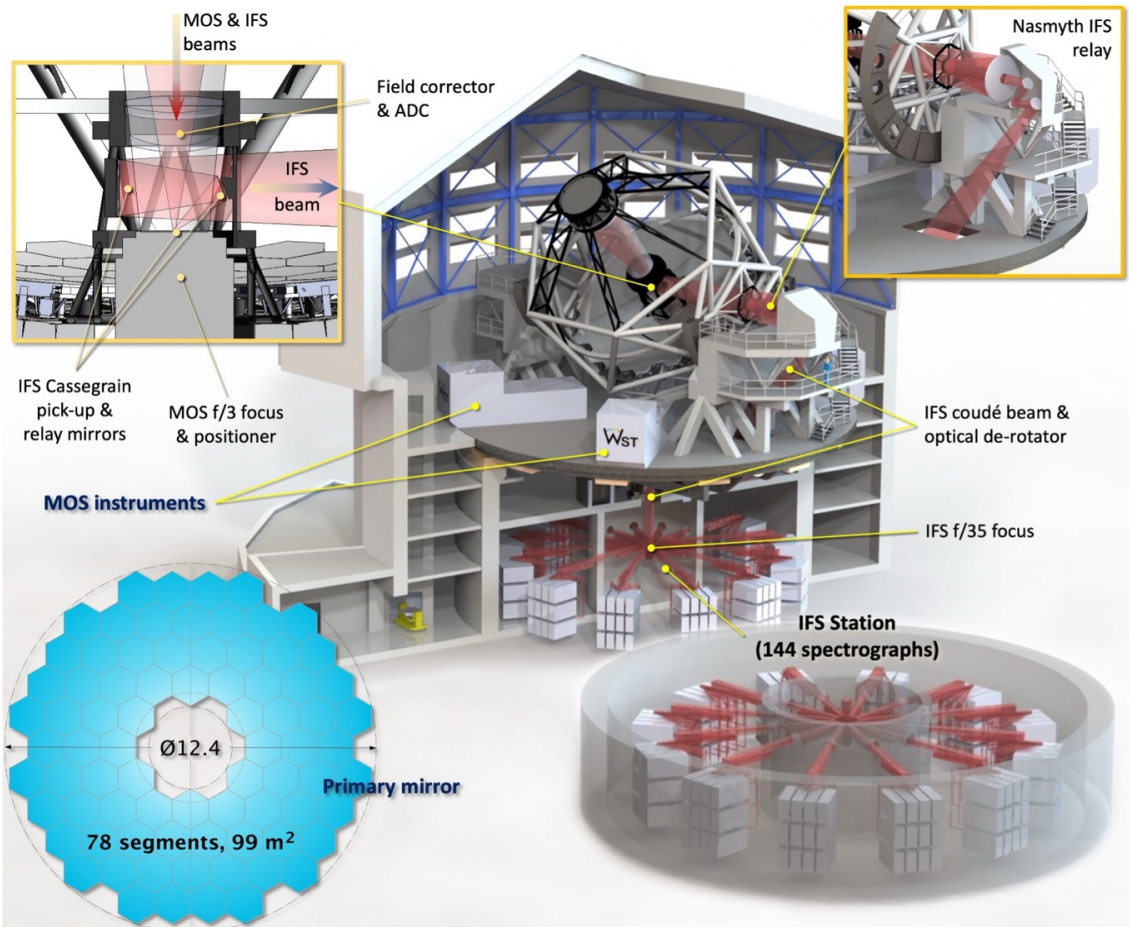

*Figure 17: WST current Facility design. The MOS spectrographs are located on the azimuth floor, for minimal fibre length. The gravity-invariant IFS station is located in the pier, below the optical de-rotator. IFS sub-field extraction (3 x 3 arcmin$^2$ out of the 13 arcmin diameter patrol field) is performed at the optical exit of the Nasmyth relay (from Bacon et al. 2024).*

The current concept design of the facility is presented in Figure 17. Note the IFS station with its 144 double-channel modules, which are needed to cover such a wide field and large simultaneous spectral range. Such a large instrument would not fit in any existing 10-meter class telescopes, demonstrating the need to design the entire system as an integrated facility.

By combining these two instruments, we have designed a highly innovative spectroscopic survey facility capable of addressing outstanding scientific questions in the areas of cosmology; galaxy assembly, evolution, and enrichment, including our own Milky Way; the origin of stars and planets; and time domain and multi-messenger astrophysics. WST's uniquely rich dataset may yield unforeseen discoveries in many of these areas. A detailed science case is presented in a white paper (Mainieri et al. 2024). WST will be proposed as the next ESO facility. If it is accepted, it will advance spectroscopy in general, and IFS in particular, to the next level.

# 6. Conclusions

Integral field spectroscopy is now a mature and essential technique for exploiting our telescopes. It took many years to collectively achieve this result, but we are now at a stage where we can contemplate the next step and elevate this technique to a new level.

On a personal note, these almost 40 years of developing integral field spectroscopy have been a great technical, scientific, and human adventure. I have been fortunate to work in this international environment with so many talented people from academia and industry. There are too many to name individually, but I would like to thank all of them for being part of this joyful and exciting journey.

Acknowledgment: the author would like to thank an anonymous referee for helpful comments on the manuscript.